\documentclass[manuscript]{aastex}
\usepackage{amssymb,amsmath}

\shorttitle{Coherent Structure in Solar Wind Ionic Composition Data in 2008}
\shortauthors{Edmondson et al.}

\begin{document}

\title{Coherent Structure in Solar Wind ${\rm C}^{6+}/{\rm C}^{4+}$ Ionic Composition Data During the Quiet-Sun Conditions of 2008}

\author{J. K. Edmondson\altaffilmark{1}, B. J. Lynch\altaffilmark{2}, S. T. Lepri\altaffilmark{1}, and T. H. Zurbuchen\altaffilmark{1}}
\affil{\altaffilmark{1}Department of Atmospheric, Oceanic, and Space Science, University of Michigan, Ann Arbor, MI 48109}
\affil{\altaffilmark{2}Space Sciences Laboratory, University of California, Berkeley, CA 94720}

\begin{abstract}

This analysis offers evidence of characteristic scale sizes in solar wind charge state data measured in-situ for thirteen quiet-sun Carrington rotations in 2008. Using a previously established novel methodology, we analyze the wavelet power spectrum of the charge state ratio ${\rm C}^{6+}/{\rm C}^{4+}$ measured in-situ by ACE/SWICS for 2-hour and 12-minute cadence. We construct a statistical significance level in the wavelet power spectrum to quantify the interference effects arising from filling missing data in the time series \citep{Edmondson13}, allowing extraction of significant power from the measured data to a resolution of 24 mins. We analyze each wavelet power spectra for transient coherency, and global periodicities resulting from the superposition of repeating coherent structures. From the significant wavelet power spectra, we find evidence for a general upper-limit on individual transient coherency of $\sim$10~days. We find evidence for a set of global periodicities between 4-5 hours and 35-45 days. We find evidence for the distribution of individual transient coherency scales consisting of two distinct populations. Below the $\sim$2 day time scale the distribution is reasonably approximated by an inverse power law, whereas for scales $\gtrsim$ 2 days, the distribution levels off showing discrete peaks at common coherency scales. In addition, by organizing the transient coherency scale distributions by wind type, we find these larger, common coherency scales are more prevalent and well-defined in coronal hole wind. Finally, we discuss the implications of our results for current theories of solar wind generation and describe future work for determining the relationship between the coherent structures in our ionic composition data and the structure of the coronal magnetic field.

\end{abstract}


\section{Introduction}

Owing to the rapid expansion of the solar wind, the charge states of heavy ions freeze-in relatively low in the corona \citep{Hundhausen84}. In-situ measurements of heavy ion composition ratios for elements such as carbon and oxygen reflect the coronal electron temperature and density of their source region at the freeze-in point, typically at altitudes less than approximately one solar radii above the photosphere \citep[e.g.][and references therein]{Bochsler86, Geiss95b, Landi12}. After they freeze-in, ionic charge state values remain constant in-transit throughout the heliosphere, unaffected by solar wind turbulence, stream-stream interactions, etc. Thus, the structure and variability of the charge state ratios, such as ${\rm C}^{6+}/{\rm C}^{4+}$ and ${\rm O}^{7+}/{\rm O}^{6+}$, offer insight into the dynamics of the inner-corona at freeze-in temperatures on the order of one million degrees \citep[][and references therein]{vonSteiger1997, Zhao09, Landi12, Gilbert12}. Identifiable temporal scales within the compositional variability therefore provide a direct measure of the temporal and/or spatial variability of the coronal origin of the solar wind.

In particular, transient structures such as the ubiquitous X-ray bright points, coronal jets, and discrete changes in coronal hole boundaries \citep[e.g.,][]{Bromage2000, Madjarska2009, Subramanian2010, Kahler2010}, as well as the frequent release of plasmoids at the tip of streamers \citep[e.g.,][]{Suess1996, Woo97, Wang2000}, may be reflected in the relatively short identifiable timescales observed in situ in the solar wind near 1 AU. Longer identifiable time scales might suggest larger, relatively steady-state structures associated with the global magnetic field structure and its evolution, such as quasi-stationary active regions, low-latitude coronal holes, and the quasi-rigid rotation of significant equator-ward coronal hole extensions \citep{Timothy1975, Wang1993, Insley1995, Zhao1999}. Even the spatial scales of a highly structured coronal magnetic field (such as the S-Web model of \citet{Antiochos11} and co-workers) may be identifiable in the temporal structure of solar wind ionic charge state ratios measured in situ near 1 AU. In this case, ionic charge states emanating from separate coronal environments, and therefore having distinct signatures, are measured adjacent to one another in situ.

Many studies have been performed over the past several decades to identify and characterize common solar and heliospheric timescales, ranging from 10 years to 1 hour. 
For the most part, typical identifiable timescales longer than a Carrington Rotation, on the order of 51, 77, 102, 128, and 154 days have been associated with global solar magnetic field evolution, sunspot numbers, solar flares \citep{Rieger1984, LeanBrueckner1989, CluadeGonzalezetal1993, Caneetal1998}. Ephemeral periodicities associated with solar flares (27, 59, 137, and 330 days) and coronal mass ejections (37, 97, 182, and 365 days) have been reported by \citet{Polygiannakisetal2002}. Timescales of order the Carrington Rotation, 27 days and its harmonics (e.g., 13.5, 9, and 6.7 days) have been consistently identified from solar wind speed, IMF, and geomagnetic data \citep[e.g.,][]{Bolzan05, Fenimore78, GonzalezGonzalez1987, Gonzalez93, MursulaZieger1998, Nayar01, Nayar02, SvalgaardWilcox1998}. \citet{Temmer07} also linked the 9 day timescale to coronal hole variability in the declining phase of solar cycle 23, by comparing coronal hole areas with daily averages of solar wind parameters, speed, density, temperature, and IMF magnitude. \citet{Neugebauer97} investigated polar microstreams using a wavelet analysis of one-hour average wind speed data from Ulysses, and found a favored timescale of 16 hours. Ionic composition data have typically been excluded from the timing and spectral analyses perfomed on the IMF and bulk solar wind plasma properties. \citet{Zurbuchen00} showed the ${\rm O}^{7+}/{\rm O}^{6+}$ ratio had an autocorrelation $e$-folding time of $\sim$10~hours and a spectral index of $-2$, indicative of a time series with many discontinuities. Recently, \citet{Landi12} employed a sonification analysis \citep{Alexander2011} of the ${\rm C}^{6+}/{\rm C}^{4+}$ and ${\rm O}^{7+}/{\rm O}^{6+}$ data to show the 27~day rotation period and its first 3 harmonics (13.5, 9, 6.7 days) in the resulting Fourier power spectra. Work by Thompson and co-workers, using 1 hour average interplanetary magnetic field (IMF) data, argue that long-lived signatures of gravity-mode oscillations and solar pressure modes in the range 1 - 140 $\mu$Hz (of order 10 days - 2 hour time scales) \citep{Thompson95,Thompson01} are present in interplanetary low-energy charged particles.

The objective of this investigation is to apply a Morlet wavelet transform \citep{GrossmanMorlet1984} analysis to ACE/SWICS composition time series at both 2-hour average (standard) and 12-minute average (the highest resolution investigated by SWICS), in order to identify and characterize the common timescales within solar wind composition measurements. The common understanding that has emerged from previous studies suggests that signatures of the spatial and temporal evolution of the solar (coronal) magnetic field structure and its dynamics are reflected in the structure of and evolution of the solar wind. We choose 2008 as a period of relatively low activity during solar minimum in an attempt to minimize the influence of large scale transients (ICMEs), e.g., \citet{Kilpua2009} report only 9 ICMEs observed at L1 during 2008. While the coronal magnetic field is usually in its simplest global configuration at solar minimum, we note that the cycle 23 minimum was unusual. This time period does not exclude the formation and evolution of active regions and mid-latitude coronal holes; in fact, even the most cursory search of this time period shows plenty of these types of structures.

The wavelet analysis and the methodology of constructing the power spectra confidence levels are described in detail by \citet{Edmondson13}, however we will briefly summarize aspects of the procedure as needed. The layout of this paper is as follows. Section~\ref{S:NoMeasurementGaps} characterizes the statistical properties of the data gaps in the ACE/SWICS data sets. In section~\ref{S:Results}, we describe the results of the analysis program applied to the full year of data for both 2-hour and 12-min averages of 2008 ACE/SWICS ${\rm C}^{6+}/{\rm C}^{4+}$ data. Finally,
in section~\ref{S:Discussion} we discuss the implications of these results interpreted as physical constraints on various magnetic field structure and solar wind generation models.

\section{Data Analysis and Filling the Data Gaps}
\label{S:NoMeasurementGaps}

\begin{figure*}
\epsscale{1.0}
\plotone{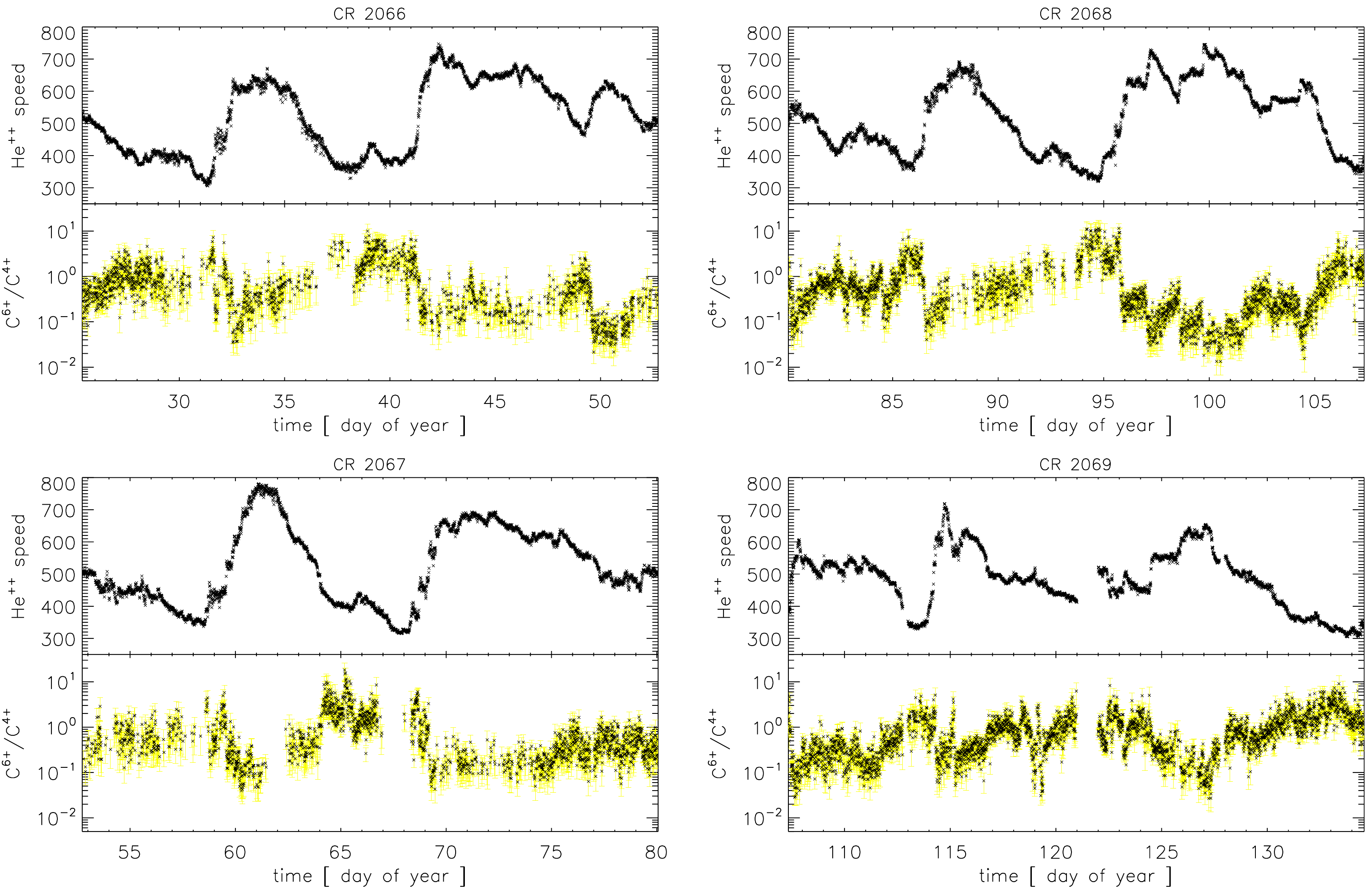}
\caption{ACE/SWICS data for four Carrington Rotations during 2008
showing large scale repeating solar wind stream structures. Each panel
plots the ${\rm He}^{2+}$ bulk speed and the ${\rm C}^{6+}/{\rm C}^{4+}$ data at 12-minute resolution. The
error bars show the 1-$\sigma$ statistical variation during the averaging
period. The intermittent data gaps in the ionic composition measurements
are clearly visible.
               \label{RepeatingStructures}}
\end{figure*}

Figure \ref{RepeatingStructures} shows the ${\rm C}^{6+}/{\rm C}^{4+}$ charge-state ratio and the solar wind (${\rm He}^{2+}$) speed which is identical to the solar wind speed to within one Alfv$\acute{\text{e}}$n speed \citep{Marsch1981,Marsch1982,Neugebauer1994} measured by ACE/SWICS over four Carrington Rotations (CR), 2066--2069, during 2008. In all four CR panels, we identify the non-coronal hole wind (NCHW) intervals, sometimes referred to ``slow solar wind'', based on criteria applied to the value of ${\rm O}^{7+}/{\rm O}^{6+} \ge 0.145$ charge state ratio \citep{Zhao09}, as well as enough event counts measured in a given duration (chosen greater than one event count per day) to differentiate between wind-type sources and statistical outlier points within a given source. These criteria identify intervals of statistically meaningful, periodically repeating, elevated charge state ratios over an approximately five day period in the middle of each CR, and whatever coronal source structure/process responsible for this repeating NCHW seems to have dissipated by CR 2069. From this repeating behavior in the four CRs of Figure \ref{RepeatingStructures}, we see clear, albeit, \textit{qualitative} evidence of a coherent structure in the solar wind composition data. Applying a wavelet analysis methodology, we can quantify all such common timescales in composition data to a resolution 24-minutes for the 12-minute cadence data set, and 4-hours for the 2-hour cadence data set.

The wavelet transform of a time series $T(t)$ is given by
\begin{equation} \label{E:WaveletTransform}
W \left( t , s \right) = \int {\rm T} ( t' ) \ \psi^* ( t' , t , s ) \ dt'.
\end{equation}
\noindent In our calculations, we use the Morlet wavelet basis. The Morlet family is a time-shifted, time-scaled, complex exponential modulated by a Gaussian envelope,
\begin{equation}
\psi \left( t' , t , s \right) = \frac{\pi^{1/4}}{\vert s \vert^{1/2}} {\rm exp}\left[ i \omega_{0} \left( \frac{t' - t}{s} \right) \right] {\rm exp}\left[ - \frac{1}{2} \left( \frac{t' - t}{s} \right)^2 \right]
\end{equation}
\noindent where $ \left( t' , t \right) \in I_{T} \times I_{T} \subset \mathbb{R} \times \mathbb{R}$ is the time and time-translation center, respectively, and $s \in I_{S} \subset \mathbb{R}$ is the time scale over which the Gaussian envelope is substantially different from zero. The $\omega_{0} \in \mathbb{R}$ is a non-dimensional frequency parameter defining the number of oscillations of the complex exponential within the Gaussian envelope; we set $\omega_{0} = 6$, yielding approximately three oscillations.

The wavelet power spectrum is given by, $\mathcal{P} \left( t , s \right) = \vert W \left( t , s \right) \vert^2$. \citet{TorrenceCompo1998} identify a bias in favor of large-scale features in the canonical power spectrum, which they attribute to the width of the wavelet filter in frequency-space; at large scales the function is highly compressed yielding sharper peaks of higher amplitude. Equivalently, high frequency peaks tend to be underestimated because the wavelet filter is broad at small scales. \citet{Liu2007} showed this effect is the difference between the energy and the integration of the energy with respect to time, and thus may be rectified, in practice, by multiplying the wavelet power spectra by the corresponding scale-frequency. Thus, throughout this paper we use the rectified power spectrum, given by, 
\begin{equation} \label{E:RecPowerSpectrum}
\mathcal{P} \left( t , s \right) = \vert s \vert^{-1} \vert W \left( t , s \right) \vert^2
\end{equation}

The wavelet transform, and subsequent (rectified) power spectrum, equations (\ref{E:WaveletTransform}) and (\ref{E:RecPowerSpectrum}), require a complete time series and, as seen in Figure~\ref{RepeatingStructures}, we must deal with data gaps \citep[see e.g.,][Appendix A for discussion of why data gaps arise]{Thompson01}. \citet{Edmondson13} evaluated several different procedures for filling the data gaps (e.g., linear interpolation, constant mean value, constant RMS value), characterizing the filler signal power and the resulting interference power that naturally arises from the nonlinear interaction in the transform of a superposition of the good data and the filler signal. A comparison power is then used to construct an 80\% confidence level against the null hypothesis that the total power spectrum at a given correlation timescale is due to either the filler signal itself, or spurious interference introduced by the filler signal (i.e., power above the 80\% significance level only has a 20\% chance of being due to the filler signal or interference effects). The wavelet power significance level procedure allows useful information to be derived from even sparsely populated data sets. Here we apply the novel procedure described by \citet{Edmondson13} to the 2-hour and 12-minute averages of the ${\rm C}^{6+}/{\rm C}^{4+}$ ionic composition ratio data.
 
The low flux conditions in 2008 meant the SWICS instrument aboard ACE was only able to record valid 12-minute average measurements 52.76\% of the time for ${\rm C}^{6+}/{\rm C}^{4+}$. Despite the relatively low filling factor, we will show our wavelet significance procedure enables us to identify and characterize coherent structure over a wide range of correlation timescale in the 12 minute ${\rm C}^{6+}/{\rm C}^{4+}$ wavelet power spectra. Table~\ref{tb:gaps} summarizes the distribution of the data gaps in the ${\rm C}^{6+}/{\rm C}^{4+}$ 12-min and 2-hour data. In general, the distributions of gap duration and maximum gap lengths are similar across the respective carbon data sets \citep[e.g., see][]{Edmondson13}. The 2-hour averaged data are robust with 97.5\% valid measurements.

A feeling for the distribution of ${\rm C}^{6+}/{\rm C}^{4+}$ data-gaps can been obtained from the top panels in Figure~\ref{CarbonChargeStateRatio2hr} (2 hour ${\rm C}^{6+}/{\rm C}^{4+}$) and Figure~\ref{CarbonChargeStateRatio12m} (12 min ${\rm C}^{6+}/{\rm C}^{4+}$). The black points are valid measurements while the green points show the values of the data gap filler signals used in each case. \citet{Edmondson13} employed a Monte-Carlo technique to argue, for both cadence data sets, a filler signal of a simple linear interpolation yields an ensemble-averaged time-integrated wavelet power per scale (which for the Morlet basis is equivalent to the standard Fourier power spectrum), most closely matches that of an  ensemble average ideal gap-free signal.

\section{Results}
\label{S:Results}

\subsection{Characteristics of the Wavelet Power Spectra}
\label{SS:WaveletPowerCharacteristics}

\begin{figure*}
\epsscale{0.92}
\plotone{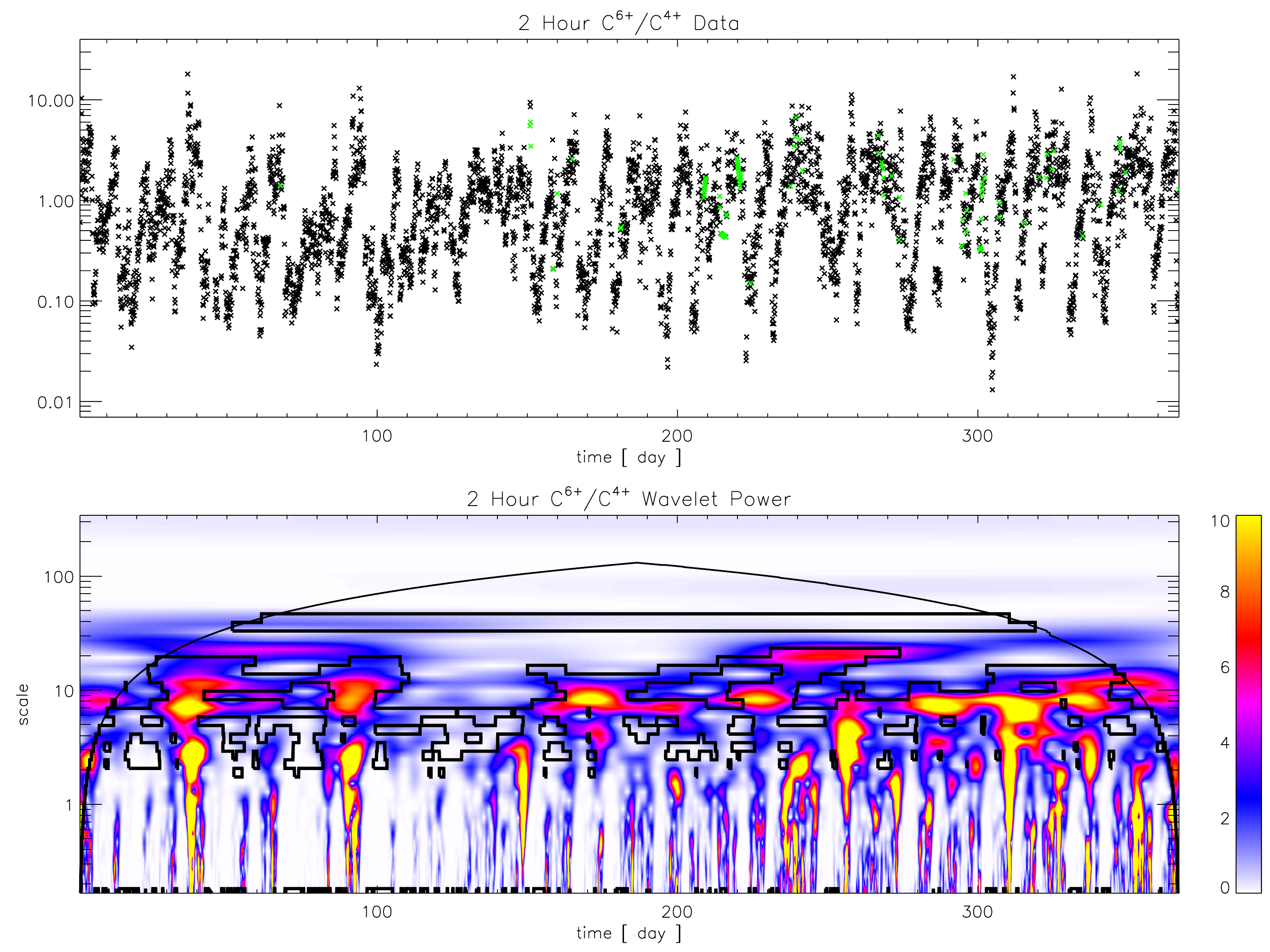}
\caption{Top panel, ACE/SWICS 2-hour average ${\rm C}^{6+}/{\rm C}^{4+}$
charge state ratio. The black data points indicate good measurements while
the green points show the filler signal applied to the data gaps. Bottom
panel, the corresponding wavelet power spectrum with contours of the
$80\%$ significance level shown in black.
               \label{CarbonChargeStateRatio2hr}}
\end{figure*}

\begin{figure}
\epsscale{1.00}
\plotone{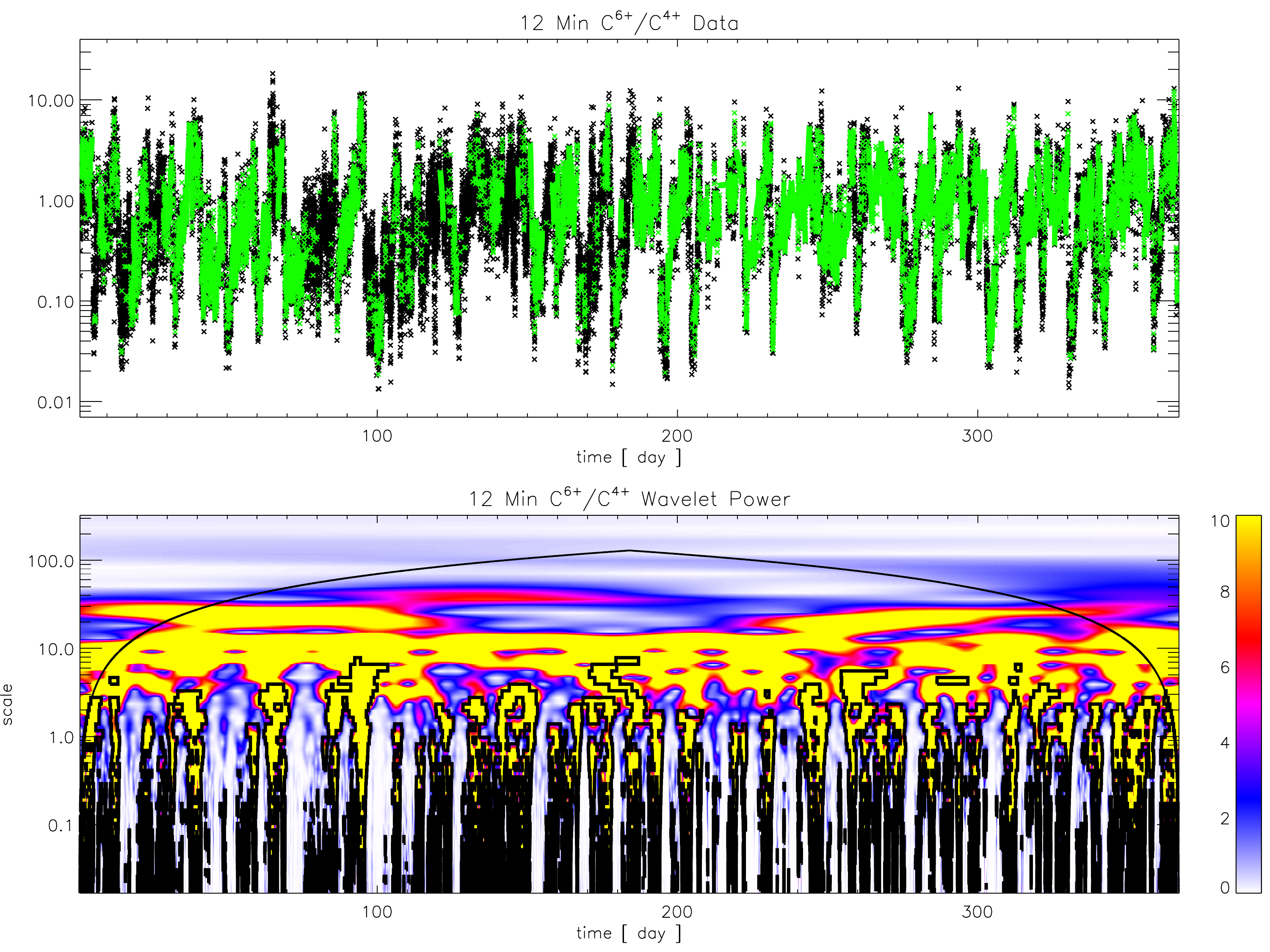}
\caption{${\rm C}^{6+}/{\rm C}^{4+}$ 12-minute data and power spectra
in the same format as Figure~\ref{CarbonChargeStateRatio2hr}.
}
               \label{CarbonChargeStateRatio12m}
\end{figure}

Figures~\ref{CarbonChargeStateRatio2hr} and \ref{CarbonChargeStateRatio12m} show the full charge state ratio time series and corresponding total wavelet power spectra for ACE/SWICS ${\rm C}^{6+}/{\rm C}^{4+}$ data at 2-hour and 12-minute averages, respectively for 2008. Our analysis focuses on ${\rm C}^{6+}/{\rm C}^{4+}$ data because this particular ratio exhibits superior statistics over other comparable charge states, as well as the fact that the ionic charge state freezes-in very close to the Sun in a nearly local fashion \citep{Landi12}. The domain of analysis in the wavelet power spectrum is defined by two boundaries: the cone of influence due to the padding of zeros beyond the boundaries of the time series required to keep the wavelet transform well defined throughout the entire domain of interest \citep{TorrenceCompo1998}, and the 80\% significance level against the interference of the data gap filler signal \citep[]{Lachowicz09, Edmondson13}. The former quantifies the influence of the zeros padding in the wavelet power spectra. The latter standard quantifies the wavelet power spectrum effects due to the filling-in of data gaps with particular values. 

The most obvious characteristic exhibited in the power spectra patterns (lower panels of Figures~\ref{CarbonChargeStateRatio2hr}, \ref{CarbonChargeStateRatio12m}) is that in both cases, the strongest relative significant power structures occur in a timescale band between approximately 1 and 30 days, with increasingly localized strong power extensions to smaller correlation scales throughout the year. \textit{Nearly} all power structures are significant at scales below approximately 2 days for both data set cadences. In fact, all 2-hour data set power is significant at scales between approx. 2 days and the Nyquist scale. This effect reflects the ability of the significance level procedure to identify significant power at the small scales above the effects of the filler signal and its interference for sparsely populated measurement intervals.

On the other hand, few significant power structures occur at large timescales in both cases. For the 2-hour data set, save a significant power band between 30 - 50 days (most likely reflecting the evolution of Carrington rotation timescale patterns as they evolve from the Sun to 1 AU), there is no significant power at timescales greater than approx. 20 - 30 days. In the case of the 12-minute data set, there is no significant power at timescales greater than approx. 10 days. Such large-scale effects reflect the significance procedure rejection due to interference between the measured data and the filler signal.

A band of strong significant wavelet power patterns in the 2-hour data (Figure \ref{CarbonChargeStateRatio2hr}) extend about timescales on the order of 10 days, with some evidence of larger timescales between 10 and 30 days. This effect is present to a much lesser extent in the 12-minute cadence wavelet spectra as well; though due to the amount of filler signal required, very little of the power on this timescale is identified as significant. For reasons laid out in the rest of this paper, we interpret this order 10 day scale as demarcating the upper-limit on coherency timescale of individual coherent (pulse-like) structures, and the significant power patterns at larger scales reflect the superposition of the wavelet power associated with a series of repeating structures. \citet{Edmondson13} discussed the wavelet response to a pulse-like structure in their Appendix A; relatively localized enhancements expand to larger correlation scales, spreading out forwards and backwards in time. Power at the scale of the periodicity of the pulses (often greater than their actual width) is obtained through the superposition of this wavelet power diffusion to longer correlation scales that now overlap in time. 

This (roughly) 10-day upper-limit can be \textit{qualitatively} identified from the patterns in the full time series of carbon charge state ratio values (top rows of Figures \ref{CarbonChargeStateRatio2hr}, \ref{CarbonChargeStateRatio12m}). In both cases, the duration between local maximum values occurs at a cadence of order approximately 10 days. Thus, the duration of coherent individual pulse-like structures in the charge state values are, in turn, of order 10 days. In addition, the significant power structures at scales greater than 10 days can, in both cases, be \textit{qualitatively} identified with repeating coherent pulse-like structures of order 10 days in the charge state ratio values.

As an example, we point to the clear set of three strong power structures in both data sets that occur around times $t$ = 35, 65, and 95 days. The significant wavelet power pattern associated with these structures clearly exhibits coherency at all scales below 10 days in the 2-hour data, and 5 days in the 12-min data, suggesting they are produced by a superposition of even smaller scale structures. In both power spectra, there is strong, though not significant, power on the order of 20 - 30 days corresponding to the source repeating at the Carrington rotation time scale. Due to filler signal interference effects, however, this strong power at the Carrington rotation timescale was not identified as significant to 80\% confidence.

\subsection{Integrated Wavelet Power per Scale: Global Periodicities}
\label{SS:GlobalOscillationFrequencies}

\begin{figure}
\epsscale{0.7}
\plotone{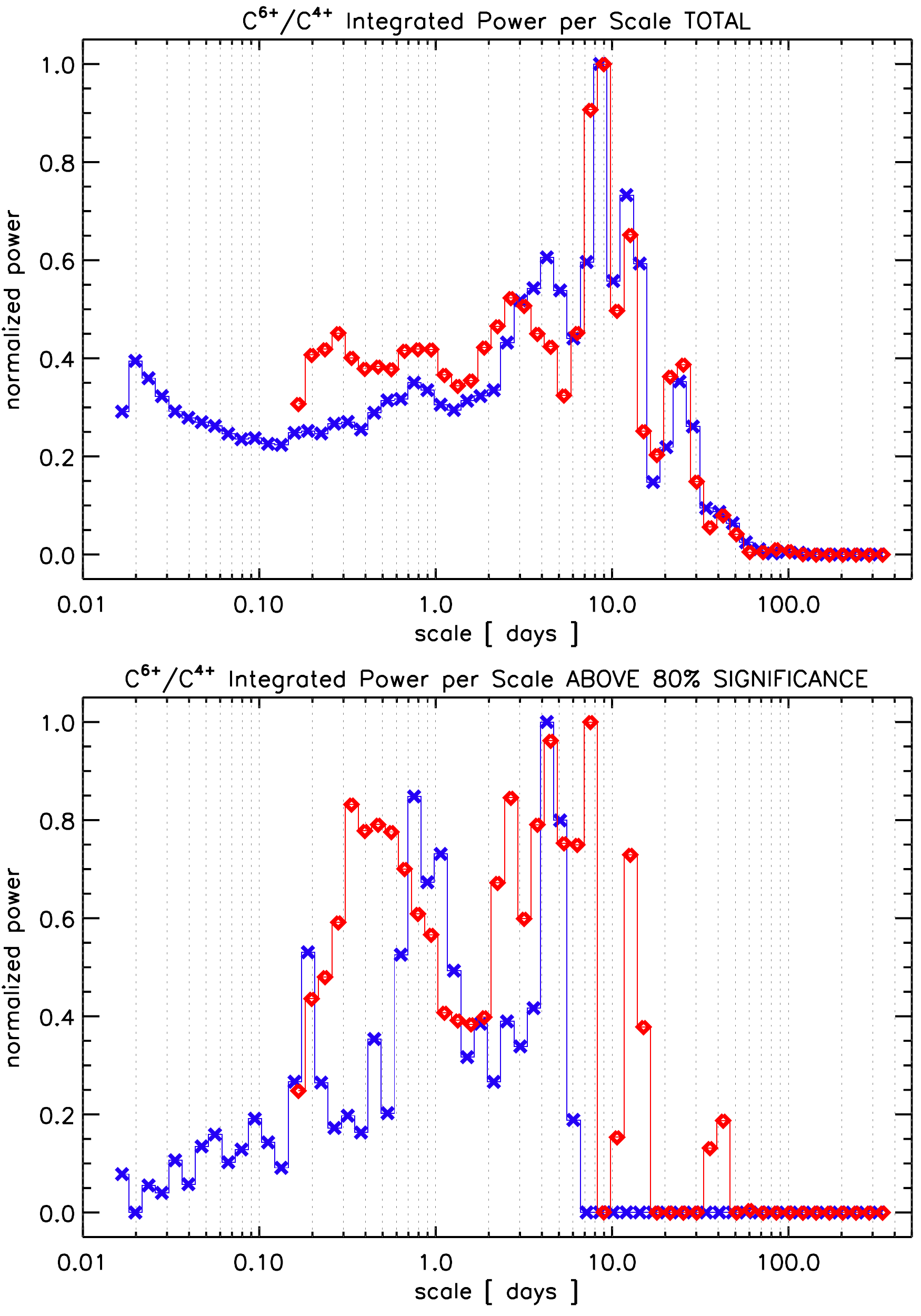}
\caption{Comparison of the normalized integrated power per scale for the ${\rm C}^{6+}/{\rm C}^{4+}$ data: Top panel plots the integrated \textit{total} wavelet power; bottom panel plots the integrated \textit{significant} wavelet power. The 12-minute data are blue, 2-hour data are red. Note, the similar timescale neighborhoods corresponding to the local maxima in the respective curves; the differences in the respective amplitudes are due to the particular filler signal model and normalization procedure, and specific differences in the 80\% significance level contours.}
               \label{IntegratedPowerPerScale}
\end{figure}

We quantify the global periodicities within the data sets by integrating the significant power in the wavelet spectra for each correlation scale over time, known as the global wavelet power spectrum. For the Morlet basis, this is equivalent to a Fourier decomposition of the time series. Figure~\ref{IntegratedPowerPerScale} shows the integrated power per scale for the different cadences plotted in different colors (2-hour averages shown in red, and 12-minute averages shown in blue). 

The top panel of Figure~\ref{IntegratedPowerPerScale} shows the integrated power per scale of the \textit{total} wavelet power bounded only by the cone of influence, therefore the integrated power per scale spectrum includes the superposition of power associated with the valid ${\rm C}^{6+}/{\rm C}^{4+}$ measurements, as well as the filler signal applied to the data gaps, and the resulting nonlinear interference pattern. Note the agreement in the integrated power spectra between the respective cadences at timescales above about 0.5 - 1 days. Below this timescale, the two curves diverge due to filler signal and normalization effects, however the important point is that local maxima of the respective curves all fall within similar timescale vicinities. 

The bottom panel of Figure~\ref{IntegratedPowerPerScale} plots the integrated power per scale of only the \textit{significant} wavelet power, i.e., the values that exceed the 80\% confidence level indicated by regions within the contours plotted in Figures~\ref{CarbonChargeStateRatio2hr} and \ref{CarbonChargeStateRatio12m}. In this case, the differences in the curve amplitudes are not only due to the particulars of the filler signal model and normalization procedure, but also specific differences in the 80\% significance level contours. However, again, the respective timescales corresponding to the curve local maxima that remain all occur within similar neighborhoods. There is little difference in the trends of the curves between the 2-hour spectra of the top and bottom panels -- as expected due to the very small percentage of missing data ($\sim$2.5\%). On the other hand, the 12-minute spectra shows significant changes once we have applied the \citet{Edmondson13} significant power procedure to remove the effects of filling the data-gaps and the interference contribution. Specifically, no significant power above the 7 day timescale is identified with 80\% confidence that the particular structure is not the result of the filler signal or interference effects. Additionally, the 12-minute integrated power per scale spectrum peaks at small timescales (below approx. 1 day) become far more pronounced.

A detailed comparison across the significant power curves (bottom panel) suggests evidence for a set of common periodicities within the 1 - 10 day decade. Specifically, we identify three strong local peaks in the 2-hour curve corresponding to repeating harmonic timescales of 2 - 3 days, and 4 - 5 days, and 7 - 8 days. The strongest local peak in the 12-min data curve matches the 4 - 5 day timescale. However, there is evidence for the 2 - 3 day harmonic in the 12-min curve. We note, similar peaks within this same decade are identifiable in both total integrated power per scale curves of the top panel.

The lowest-frequency harmonics (above the 10 day timescale) for which we find evidence come from the 2-hour data significant power: timescales between 10 - 17 days, and 35 - 45 days. Such peaks are identifiable to a lesser extent in the total integrated power per scale. We note the strong peak between 20 and 30 days in the total integrated power (top panel) for both cadence data sets is completely annihilated by the significance procedure. The longest timescale (35 - 45 days) is most likely reflective of the Carrington rotation timescale from coronal source regions at higher than equatorial latitudes. However, regardless of significant power, in general the cone of influence boundary in the power spectra, required by the wavelet transform algorithm, severely limits the identification of these long time scales relative to a single year data set. To elucidate any physical harmonics at these very-long time scales requires data sets much longer than a single year for analysis \citep[e.g.,][]{Temmer07,Katsavrias2012}. \citet{Landi12} showed the 27~day rotation rate was identifiable in the Fourier analysis of ${\rm C}^{6+}/{\rm C}^{4+}$ and ${\rm O}^{7+}/{\rm O}^{6+}$ over a time period that included our 2008 data. Our significant wavelet power is consistent with their finding that the 43 $\mu$Hz (approximately 1/27~day$^{-1}$) frequency was more visible in carbon than oxygen, yet by the first harmonic of 85 $\mu$Hz (approximately 1/13.5~days$^{-1}$) the peaks were comparable.

Evidence for harmonic timescales below approx. 1 day in the significant integrated power per scale curves become less consistent between the data sets. The significant 2-hour curve exhibits a single coherent structure in the decade below the 1 day timescale, with two slight local peaks occurring between 7 - 8 hours (0.3 - 0.35 days), and 10 - 12 hours (0.45 - 0.5 days). On the other hand, only the two weakest peaks of the five identifiable local maxima in the significant 12-min curve in the timescale decade between 0.1 and 1 days match the above listed 2-hour small timescales. The other three identifiable peaks correspond to timescales of approx. 4 - 5 hours (0.18 - 0.2 days), 17 - 20 hours (0.7 - 0.8 days), and slightly longer than 24 hours (1 day).

Below the 4-hour Nyquist scale of the 2-hour data, there is little-to-no discernible pattern (identifiable peaks) of common oscillation frequencies in the 12 minute data. Several local peaks may be identified at approximately: 2.1 - 2.4 hours, 1.2 - 1.4 hours, 40 - 50 mins, 36 mins, and 24 mins. Even though significant power is recovered in small correlation timescales ($\lesssim$ 4 hours), this overall trend at the smallest scales suggests an effective lower-bound for the range of harmonic timescales in which the superposition of wavelet peaks arising from repeating individual coherent structures blend together; i.e., in Figures~\ref{CarbonChargeStateRatio2hr}, \ref{CarbonChargeStateRatio12m}, the (significant) wavelet power spectra exhibit a highly collimated striation pattern. The integrated power per scale thus results in a ``broad-band" response for correlation timescales $s \lesssim 4$ hours.

\subsection{Distribution of Wavelet Peaks: Characteristics of Coherent Structure}
\label{SS:SmallScaleCoherentStructureCharacteristics}

To obtain information about the range of coherency scales of transient structures \citep[see][Appendix A for discussion]{Edmondson13} in the charge state ratios, we construct a probability distribution (PDF) from the histogram of the local 2D maxima in each of the wavelet power spectra as a function of correlation scale. Here, we identify the local 2D maximum in the wavelet power within the 80\% significance contours (to illustrate, note the local power maximum - bright yellow - in Figure \ref{CarbonChargeStateRatio2hr} occurring at the approximate time and scale coordinates, $t \sim 40$ days and $s \sim 9$ days, respectively). If the neighboring points in every direction have a lower power than the point being evaluated, that peak is cataloged. We note, this definition includes maxima in the wavelet power resulting from the superposition of repeating coherent structures. Unlike the integrated power per scale, this procedure does not contain information about the absolute magnitudes of the local 2D maxima in the power spectrum, only that a local maximum exists and is significant relative to the constructed 80\% confidence level. The histogram of all significant power peaks thus gives information about the distribution (i.e., frequency of occurrence) of coherent timescales of transient structures as well as the set of correlation timescales associated with their superposition. From Figures~\ref{CarbonChargeStateRatio2hr}, \ref{CarbonChargeStateRatio12m}, local maxima in the power spectrum at the largest scales ($s \gtrsim 30$~days) are clearly the result of superposition interference effects between repeating structures, as opposed to single very-large coherent scales. For intermediate scales, ($1 \lesssim s \lesssim 10$~days), significant local maxima are like the result of both relatively individual coherent structures and some superposition of the overlap, and for the smallest scales ($s \lesssim 1$~day), local maxima are very likely to arise purely from individual coherent structures.

\begin{figure}
\epsscale{0.7}
\plotone{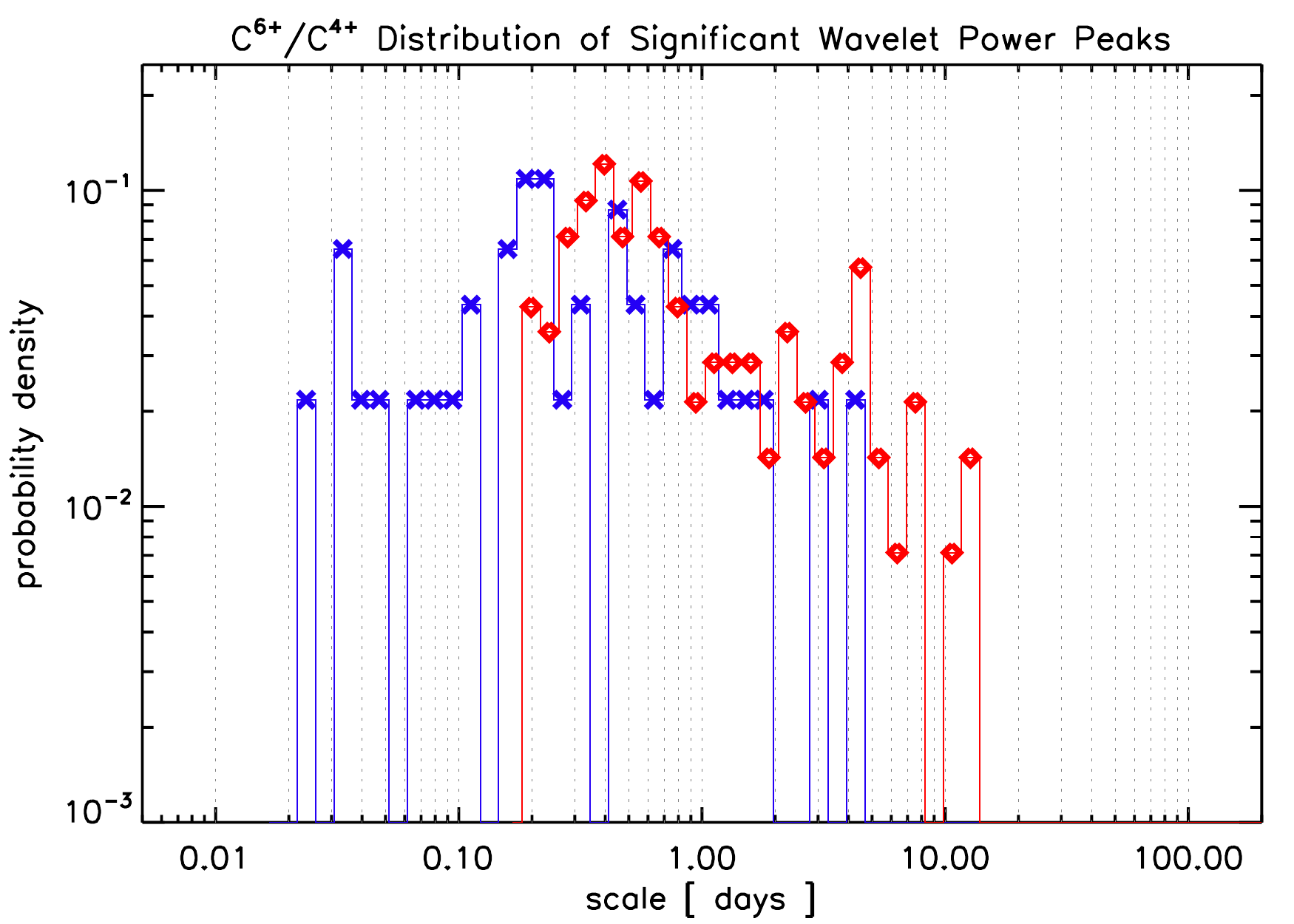}
\caption{Probability distribution of significant correlation scale rates
for the ${\rm C}^{6+}/{\rm C}^{4+}$ data (12-minute: blue; 2-hour: red).
               \label{ScaleProbabilityDistribution}}
\end{figure}

Figure~\ref{ScaleProbabilityDistribution} shows the probability distribution of the coherent structure correlation scales (i.e., rate of occurrence of local 2D maxima at a given scale) in the wavelet power spectra for the ${\rm C}^{6+}/{\rm C}^{4+}$ data. The cadence is again identified by the color: 2-hour in red, 12-minute in blue.

There are a number of interesting features in the distribution of significant power peaks of Figure~\ref{ScaleProbabilityDistribution}. First, both cadence data sets exhibit similar distributed increases in the rates of occurrence for coherency timescales in 0.1 - 1 day decade. The absolute maxima for each PDF occur in the same neighborhood, approximately 0.4 days and 0.2 days, for the 2-hour and 12-min cadence sets, respectively. Second, both distributions appear to be composed of ``dual populations": an elevated distribution superposed on a relatively constant distribution. The elevated portion of the distribution is characterized by a relatively sharp increase in the 0.1 - 0.4 day interval followed by a power-law decrease through approx. 2 day coherency scale. We note, in the large coherency scale range consisting of a constant distribution, the 2-hour PDF displays considerably more variation, compared with the 12-min distribution. However, the constant distribution seems reasonable at least through 4 - 5 day coherency timescales, where the (significant) 12 min coherency scale distribution drops to zero. Third, the local peak in the 12-min data set identifying a coherency timescale of approximately 0.03 days (45 mins) is significant enough relative to the surrounding distribution to suggest this is the smallest coherent timescale of transient structures within the data.

\subsection{Comparison With Monte Carlo Simulations of Synthetic Composition Data}
\label{SS:Comparison}

In order to underscore the physical origins of structure within the ${\rm C}^{6+}/{\rm C}^{4+}$ ionic composition data, we have examined the wavelet properties of a Monte Carlo ensemble of simulations synthetic ${\rm C}^{6+}/{\rm C}^{4+}$ data using the method outlined in \citet{Edmondson13} based on the first-order Markov chain model used by \citet{Zurbuchen00}. We have calculated 100 realizations of the model carbon time series at both the 2 hour and 12 minute cadences. The model time series is generated by
\begin{equation}
Z_n = Z_{n-1}{\rm exp}\left[ -\Delta t/\tau_{1/e} \right] + X_n,
\end{equation}
\begin{equation}
Y_n = {\rm exp}[ \sigma_\ell \hat{Z_n} + \mu_\ell ],
\end{equation}
where the amount of ``memory" in each step of the Markov process $Z_n$ is controlled by the exponential decay term ${\rm exp}\left[ -\Delta t/\tau_{1/e} \right]$ and $X_n$ is an independent random number drawn from a normalized Gaussian distribution. Here we have taken an $e$-folding time of $\tau_{1/e} = 0.42$~days (10 hours) and the $\Delta t$ term represents the temporal cadence of the particular time series modeled: $\Delta t = 0.083$~days (2 hour) and $\Delta t = 0.0083$~days (12 minute). Therefore, ${\rm exp}\left[ -\Delta t/\tau_{1/e} \right] $ gives values of 0.81 (0.98) for the 2-hour (12-minute) data. The model ${\rm C}^{6+}/{\rm C}^{4+}$ values are given by $Y_n$ which depends on $Z_n$ normalized to unit variance ($\hat{Z_n}$) and tuned to the particular time series of interest with two parameters representing the mean ($\mu_\ell$) and standard deviation ($\sigma_\ell$) of the logarithm of the ${\rm C}^{6+}/{\rm C}^{4+}$ data. The 2-hour data give ($\mu_\ell$, $\sigma_\ell$) = ($-0.15$, $0.52$) and the 12-minute data ($\mu_\ell$, $\sigma_\ell$) = ($-0.26$, $0.52$).

\begin{figure*}
\epsscale{1.0}
\plotone{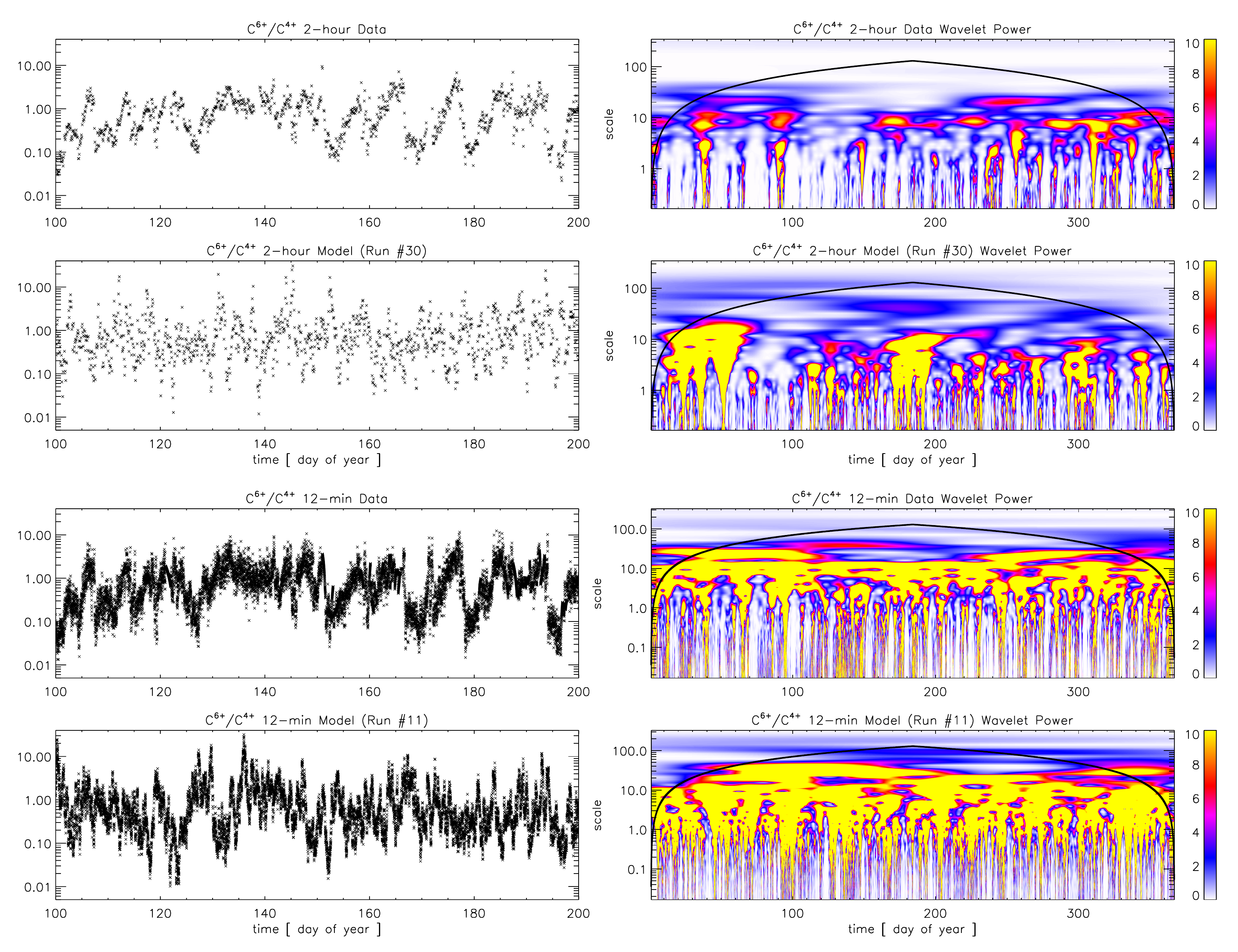}
\caption{Comparison of the ${\rm C}^{6+}/{\rm C}^{4+}$ data with two realizations of the Zurbuchen et al. inspired first-order Markov process. Left column shows 100~days of the 2-hour data, a 2-hour model run, the 12-minute data, and a 12-minute model run. The right column shows the respective wavelet power for the entire 13 Carrington Rotation duration. Each of the power spectra are plotted with a common color scale range to show the qualitative, systematic differences between the data and the model runs. 
               \label{figA1}}
\end{figure*}

Figure~\ref{figA1}, left column, compares a 100-day interval of the observed ${\rm C}^{6+}/{\rm C}^{4+}$ data with one model realization in the ensemble set at each temporal cadence. The top two panels show the 2-hour data and the 2-hour model run \#30 whereas the bottom two panels show the 12-minute data and 12-minute model run \#11. In the observations, the data gaps have been filled via the methods described in Section~\ref{S:NoMeasurementGaps} for each case.  The synthetic model time series reproduce the overall mean and range of values of the observations, as they should given the parameter selection. However, in both the 2-hour and 12-minute cases, the model runs show obvious differences. There is more high frequency scatter in the model runs than present in the data and therefore, the models show less ``coherency" in intermediate scale features (e.g., the $\sim$3--5 day width of enhancements). This is easily visible in the 2-hour comparison but also present in the 12-minute comparison. 

The differences between the observations and synthetic time series are also readily apparent in the wavelet power spectra. The right column of Figure~\ref{figA1} show the comparison between the full year's wavelet power spectra for the 2-hour gap-filled observations and model run \#30 (top two panels) and the 12-minute observations and mode run \#11. In all cases we have used the same color scale ranges for the power level, saturating the maximum values at 10 to highlight the overall differences in the respective power spectra. Both the 2-hour and 12-minute Monte Carlo runs have much higher wavelet power, compared with the measured data time series, arising from the both the higher frequency scatter in the model time series, as well as the non-linear interference wavelet response pattern generated by the superposition of successive of coherent pulses \citep[e.g., see][]{Edmondson13}. Qualitatively, each Monte Carlo run power spectra has maximum values roughly an order of magnitude higher than the corresponding data. 

\begin{figure*}
\epsscale{1.0}
\plotone{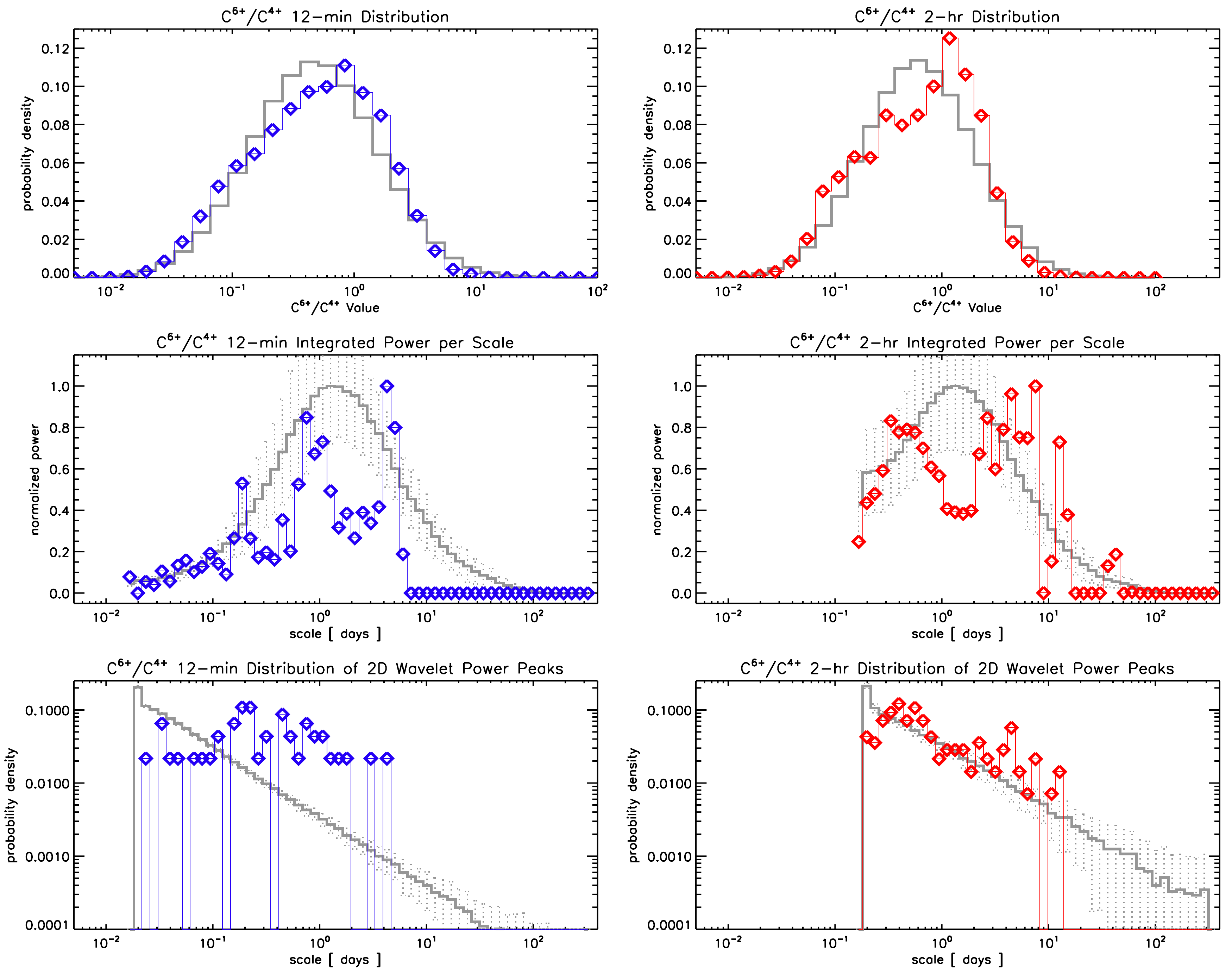}
\caption{Comparison of the statistical and spectral properties of the ${\rm C}^{6+}/{\rm C}^{4+}$ observations with the Monte Carlo ensemble averages for the 12-minute data (left column) and 2-hour data (right column). The top row plots the distribution of values with the modeling results as the gray line, the data as blue and red points. The middle row plots the integrated power per scale; the data spectra are the significant integrated power spectra above 80\% confidence level from the lower panel of Figure~\ref{IntegratedPowerPerScale}, whereas all the wavelet power in the model ensemble can be considered ``significant". The bottom row plots the distribution of 2D peaks in the wavelet power spectra. 
               \label{figA2}}
\end{figure*}

Figure~\ref{figA2} shows a comparison of the statistical and spectral ensemble-averaged properties of the modeled charge state data with the corresponding data properties. The top row shows the total probability distribution histogram of the ${\rm C}^{6+}/{\rm C}^{4+}$ values for the 12 minute averages (left column, blue) and 2 hour averages (right column, red), with the ensemble modeling results shown as the thick gray line. The middle row plots the ensemble-averaged normalized integrated power per scale and the normalized integrated power per scale above the 80\% significance level of the data (from section~\ref{SS:GlobalOscillationFrequencies}). The bottom row plots the ensemble-averaged probability distribution histogram of the number of 2D peaks and the distribution of significant peaks in the data (from section~\ref{SS:SmallScaleCoherentStructureCharacteristics}). 

The similarities and differences between the Markov process modeling of the ${\rm C}^{6+}/{\rm C}^{4+}$ ratio and the actual data are also apparent in Figure~\ref{figA2}. While the distribution of the actual model and data values are -- by construction -- quite similar, the integrated power per scale plots show some fundamentally different characteristics. The most important difference is that the in-situ data contain spatial and/or temporal signatures of the physical source region and generation mechanisms for the ${\rm C}^{6+}/{\rm C}^{4+}$ ionic composition, while the Monte Carlo ensemble-averages, on the other hand, have no such corresponding structures. As a consequence, the integrated power per scale spectra of the model set averages are smooth, gaussian-like shapes centered at the 1 - 2 day timescale, while the corresponding curves for the data sets exhibit (several) distinct global oscillation frequencies. In addition, the distribution of 2D maxima in the wavelet power as a function of timescale, corresponding to the frequency of occurrence of coherency timescales of the transient structures, in the model ensemble-average present a power-law dependence on coherency scale size ($\sim$$s^{-1}$). The corresponding 2D maxima distribution for the 12-min data set does not on the whole exhibit this behavior; the 2-hour distribution may follow a similar power-law dependence, at least up until the $s \gtrsim 1$ day scale. Both Figures~\ref{figA1} and \ref{figA2} show that, while the first-order Markov process modeling can reproduce certain statistical properties of the ${\rm C}^{6+}/{\rm C}^{4+}$ ionic composition observations, the data contain additional physical information in the shape of the wavelet power spectra and in the moderate-to-large scale range of the distribution of correlation scales unaccounted for in this modeling approach.

\subsection{Coherent Structure Scales by Solar Wind Type}
\label{SS:CoherentStructureCharacteristicsbyWindType}

\begin{figure*}
\epsscale{1.0}
\plotone{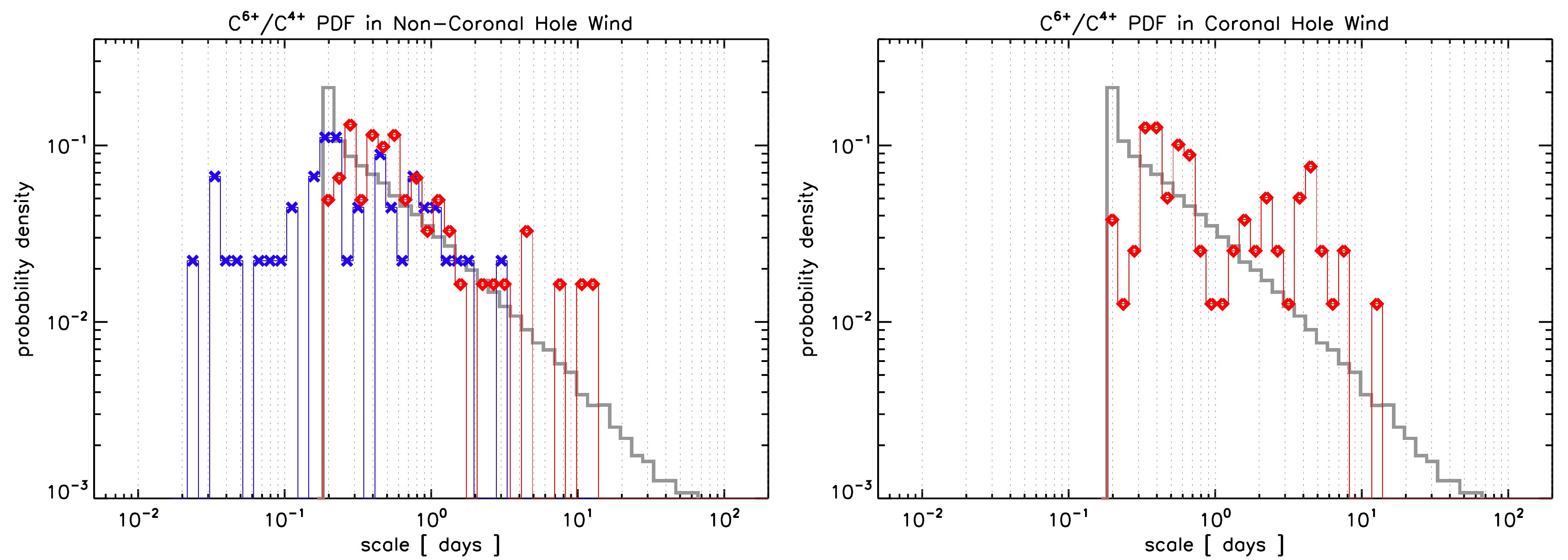}
\caption{Probability distribution of significant correlation scale rates
for 12-minute cadence (blue) and 2-hour cadence (red) ${\rm C}^{6+}/{\rm C}^{4+}$ charge state ratio data series by solar wind type. Left column: Non-coronal hole wind. Right column: Coronal hole wind. In the top row, the thick gray lines plot the Monte Carlo modeling ensemble-average distributions.  The bottom row plots the ratio of the data PDF by wind type to the Monte Carlo ensemble-average PDF.
The thick gray solid (dashed) lines show the normalized 1-$\sigma$ (2-$\sigma$) variation in the model ensemble distribution. 
               \label{ScaleProbabilityDistributionbyWindType}}
\end{figure*}

We decompose the probability distributions of correlation timescales presented in section~\ref{SS:SmallScaleCoherentStructureCharacteristics} according to wind type defined by the \citet{Zhao09} conditions evaluated on the ${\rm O}^{7+}/{\rm O}^{6+}$ density ratio. Figure \ref{ScaleProbabilityDistributionbyWindType} plots the distributions of coherency correlation timescales for non-coronal hole wind (NCHW: ${\rm O}^{7+}/{\rm O}^{6+} \ge 0.145$) in the left column, coronal hole wind (CHW: ${\rm O}^{7+}/{\rm O}^{6+} < 0.145$) in the right column. Again, the 12-minute data are shown in blue, the 2-hour data are shown in red. In addition, we show the ensemble-averaged 2D peak maxima PDF derived from the Monte Carlo simulations taken from 2-hour cadence in grey. As discussed in section \ref{SS:Comparison}, the Markov process model data yield an $\sim$$s^{-1}$ power-law dependence on the correlation scale. In general, the non-coronal hole wind distributions exhibit the dual-population behavior as discussed in section \ref{SS:SmallScaleCoherentStructureCharacteristics}, agreeing with the $\sim$$s^{-1}$ power-law decrease in coherency timescales over the decade between approx. 0.2 and 2 days, and a constant distribution outside this decade with two possible physically significant exceptions around approx. 0.03 - 0.04 days (12-min data) and approx. 4 - 5 days (2-hour data). The coronal hole wind, intervals in which no significant power in the 12-min data occurs, shows substantially more variation in the distribution. The inverse power-law \textit{may} occur in the same 0.2 - 2 day decade, however, any possible agreement is for the most part lost at coherency timescales above 2 days.

Both wind types exhibit a departure from the power-law dependence (model distribution) at larger scales, the break occurring around 2 days; though this transition is much less obvious in the 2-hour coronal hole wind data. While there is evidence of NCHW correlation scale enhancements at $s = \{$ 0.3, 0.4 - 0.6, 0.7 - 0.8, 2, 3 - 4 $\}$ days, the CHW coherency scales at $s = \{$ 0.3 - 0.4, 0.6 - 0.7, 1.5, 2.5, 3 - 4 $\}$ days represent significant departures from the Markov process modeling; all distributions fall to zero for coherency scales above $\sim$10 days. Thus, we conclude that the majority of correlation scales greater than approx. 2 days are associated with coronal hole wind and therefore reflect, in some way, the larger scale coronal structure of its source region. Likewise, for correlation scales $s \lesssim 2$ days, the probability distribution of 2D wavelet power peaks in both the 12-minute and 2-hour data are, to a reasonable approximation, consistent with the distribution resulting from the Markov process model ensemble. At these correlation scales, our results confirm the stochastic nature of the solar wind origin, or at least the stochastic nature of the relevant solar wind properties that generate the ${\rm C}^{6+}/{\rm C}^{4+}$ ionic composition ratio.

\section{Discussion}
\label{S:Discussion}

This analysis offers evidence of characteristic scale sizes in ${\rm C}^{6+}/{\rm C}^{4+}$ solar wind charge state data measured in-situ for thirteen quiet-sun Carrington rotations in 2008. We analyze the wavelet power spectrum patterns of the 2-hour average and 12-minute average data cadence for global periodicities and individual coherency structures organized by wind type. These particular charge state ratios reflect coronal sources for plasma environments at temperatures of order 1~MK. Having constructed the \citet{Edmondson13} wavelet power significance levels, we are able to characterize high frequency, small correlation scale features in the high resolution 12-minute charge state ratio data despite data gaps at various temporal length distributions and percentages during low flux conditions of 2008.

From the wavelet power spectra above the $80\%$ significance level, we find evidence for a general upper-limit on the duration of individual coherent structures in the charge state ratios of $\sim$10 days. Integrating the significant wavelet power over time at each correlation scale, we find strong evidence for a global set of repeating periodicities. In general, the significant power per scales in the 2-hour cadence data exhibits two relatively broad-band responses in the 0.1 - 1 day decade (peaking at 7 - 8 hours, and to a lesser extent 10 - 12 hours), and again in the 2 - 20 day decade (with peaks occurring at 2 - 3 days, and 4 - 5 days, 7 - 8 days, and 10 - 17 days). The 12-min significant power exhibit three much narrower response patterns, with peaks occurring in the range of 4 - 5 hours, 17 - 20 hours (and to a lesser extent at 24 hours), and 4 - 5 days. The strongest overlap between occurs at the 4 - 5 day timescale. Only the 2-hour pattern exhibits significant global oscillation frequency at 35 - 40 days, possibly reflecting an evolution of the Carrington rotation timescale patterns between the Sun and 1 AU. Below the 4-hour Nyquist scale of the 2-hour data, there is little-to-no discernible pattern in the 12 minute cadence data set. 

Constructing a probability distribution function for the occurrence of significant local 2D maxima in the wavelet power spectra as a function of scale, we find evidence for a dual population for the rates of occurrence of coherent structure timescales for a given cadence in each data set, consisting of a superposition of elevated coherency distribution and an approximately constant coherency distribution. In general, the break between the populations occurs at $\sim$2 day. We have examined the dual population character by separating the 2D wavelet power maxima PDFs by solar wind type, i.e. Non-Coronal Hole Wind (NCHW) and Coronal Hole Wind (CHW) based on the \citet{Zhao09} criteria and compared these PDFs to the resulting PDFs from Monte Carlo ensemble modeling of the ionic charge state time series as a stochastic process. The Markov process synthetic model ensemble yields inverse power law dependence on correlation scale which matches the 12-minute and 2-hour data distributions between 0.2 and 2 days. The departure from the power law distribution is largely associated with time periods classified as Coronal Hole Wind, although there are less prominent (but still significant) signatures also present in the NCHW distributions. This strengthens the hypothesis that moderate correlation scale features ($0.2 \lesssim s \lesssim 10$~days) arise due to a superposition of moderate-to-large scale coherent coronal structuring of the solar wind flows. Below approx. 0.2 day timescales, there is little evidence of coherency; only a single coherency scale peak in the 12-min data is observed in the approx. 45 min - 1 hour range.

We reproduce timescales on the order of the 16-hour microstream timescale identified by \citet{Neugebauer97}, as well as the 9 day timescale that \citet{Temmer07} associated with the coronal hole evolution during 2005. The latter suggests a persistence of coronal hole extensions through 2008 -- which were certainly observed. In fact, the standard mapping of PFSS field lines from the source surface ecliptic plane during this unusual solar minimum showed a more complex helmet streamer structure and more low-latitude coronal hole sources than the previous minima \citep{Riley12}. The general picture we emphasize here is, individual coherent structure timescales occur below approximately 1 - 2 days, transitioning to patterns of significant power at correlation longer time scales arising from the superposition of repeating coherent structure in the charge state ratio.

The break in coherency timescale occurrence rates at $\sim$2 days, occurring in both CHW as well as NCHW, is consistent with approximate mixing turn-over timescales of the magnetic network supergranulation pattern \citep[e.g.,][]{Frazier1970, WordenSimon1976, WangZirin1989,Thompson95} and could be consistent with the persistence of solar interior oscillation mode signatures at 1 AU, as  suggested by \citet{Thompson95,Thompson01}. Below this range, any individual process that might generate a coherent timescale seems to be smeared out, simply elevating the signatures of the entire composition data distribution, whereas above this range individual timescales of repeating structures become more apparent.

Coherent structure timescales prevalent in both CHW and NCHW above $\sim$2 days more likely reflect the structure of the open field distribution and are associated with the pattern of moderate-to-fast wind streams coming from either narrow polar coronal hole extensions or low latitude coronal holes interspersed within slow wind associated with pattern of the helmet streamer belt and pseudostreamers. The NCHW originates higher up in the corona and either directly or indirectly originates in plasma that was recently populating closed-field regions. We conjecture the distribution of the smallest correlation scales that approximate an inverse power-law likely arise from transient and episodic processes associated with the dynamics of the magnetic field structure, such as interchange reconnection \citep[e.g.,][]{Fisk99, Crooker02, Wang2004, Edmondson09, Edmondson10, Pariat09}. These signatures are certainly characteristic of stochastic-like processes resulting in time series with many discontinuities, and may be supportive of the tangled discrete flux tube picture of the solar wind recently suggested by \citet{Borovsky2008} and colleagues.

Furthermore, the timing analysis reflects variability in the coronal source regions, at least at the ``solar wind source surface'' where asymptotic wind conditions have been reached. On average, the largest individual coherency scales are correlated with the largest coronal magnetic field structures (we estimate the streamer belt crossing time to be in the 2 - 5 day range). However, to map the distribution into the low corona, below the ``source surface'' requires knowledge of the coronal heating and solar wind acceleration mechanisms. Continued investigation will be required to determine to what extent the common small coherency timescales correlate with smaller scale coronal structure (for example, coronal hole substructure), and to what extent the coronal magnetic field structure scales relate to the heating mechanisms that produce the given charge state data. Detailed analysis of the length scales within a sufficiently high-resolution model of the magnetic field structure, such as the S-Web model \citep{Antiochos11}, and correlations with remote sensing observations such as SDO AIA, may find that the distribution of coherent structure in the coronal magnetic field show similarities with the common correlation scale sizes observed in solar wind composition. Such an investigation is non-trivial and beyond the scope of the current analysis, but is important future work as it would shed light on whether the correlation scales of individual coherent structures in the charge state ratios reflect consecutive sampling of solar wind sources with many different properties, temporal variation of plasma properties within a single, long-duration source, or even reflect variations in the particular heating mechanisms.

\acknowledgments

J.K.E., S.T.L., and T.H.Z. acknowledge support from NASA LWS NNX10AQ61G
and NNX07AB99G. B.J.L acknowledge support from AFOSR YIP FA9550-11-1-0048
and NASA HTP NNX11AJ65G.



\newpage

\begin{center}
\begin{table*}[t]
{\small
 \setlength{\tabcolsep}{5pt}
 \hfill{}
 \begin{tabular}{|l||c|c|c|c|c||c|}
 \hline
 & \multicolumn{3}{|c|}{Distribution of Data Gaps} & Total \# & Max Gap & \% Good \\
 \cline{2-4}
 & \% $<$ 0.1~day & 0.1 $\le$ \% $<$ 1~day & 1~day $\le$ \% & of Gaps & Duration & Data \\
 \hline
 ${\rm C}^{6+}/{\rm C}^{4+}$ 12-min & 93.1\% & 6.7\% & 0.2\% &  4623 & 2.8~days & 52.8\% \\
 ${\rm C}^{6+}/{\rm C}^{4+}$ 2-hour & 68.6\% & 25.7\% & 3.9\% &   51 & 1.7~days & 97.5\% \\
 \hline
 \end{tabular}
}
\hfill{}
\caption{Properties of the data gaps in the ACE/SWICS ${\rm C}^{6+}/{\rm
         C}^{4+}$ data for the 12-minute
         and 2-hour averages during 2008.}
\label{tb:gaps}
\end{table*}
\end{center}

\newpage

\end{document}